\documentclass[12pt]{iopart}

\usepackage{graphicx}
\usepackage{bm}

\newcommand {\nc} {\newcommand}
\nc {\ve} [1] {\mbox{\boldmath $#1$}} \nc {\la} {\mbox{$\langle$}}
\nc {\ra} {\mbox{$\rangle$}} \nc {\beq} {\begin{eqnarray}}
\nc
{\eeqn} [1] {\label{#1} \end{eqnarray}}

\nc {\eol} {\nonumber \\} \nc {\half} {\mbox{$\frac{1}{2}$}}

\begin{document}

\title{Quantum dynamics of resonant molecule formation in waveguides}

\author{V S Melezhik$^1$ and P Schmelcher$^{2,3}$}

\address{$^1$ Bogoliubov Laboratory of Theoretical Physics, Joint Institute for Nuclear Research,
Dubna, Moscow Region 141980, Russian Federation}
\address{$^2$ Physikalisches Institut, Universit\"{a}t Heidelberg, Philosophenweg 12, 69120
Heidelberg, Germany}
\address{$^3$ Theoretische Chemie, Institut f\"{u}r Physikalische Chemie, Universit\"{a}t
Heidelberg, Im Neuenheimer Feld 229, 69120 Heidelberg, Germany}
\ead{melezhik@thsun1.jinr.ru, peter.schmelcher@pci.uni-heidelberg.de}

\begin{abstract}
We explore the quantum dynamics of heteronuclear atomic collisions in waveguides
and demonstrate the existence of a novel mechanism for
the resonant formation of polar molecules. The molecular formation probabilities can be
tuned by changing the trap frequencies characterizing the transverse
modes of the atomic species. The origin of this effect is the confinement-induced
mixing of the relative and center of mass motions in the atomic collision process
leading to a coupling of the diatomic continuum to center of mass excited molecular states in
closed transverse channels.
\end{abstract}

%Uncomment for PACS numbers title message
%\pacs{00.00, 20.00, 42.10}
% Keywords required only for MST, PB, PMB, PM, JOA, JOB?
%\vspace{2pc}
%\noindent{\it Keywords}: Article preparation, IOP journals
% Uncomment for Submitted to journal title message
%\submitto{\JPA}
% Comment out if separate title page not required

\maketitle

\section{Introduction}
Dimensionality plays a crucial role for interacting ultracold quantum gases
\cite{pethick08,pitaevskii03,grimm00,folman02}.
For a strong two-dimensional confinement, i.e. a tight waveguide, and a close to resonant
scattering process the atoms can feel the trapping potential
within an individual atom-atom scattering process. A striking manifestation herefore
is the confinement-induced resonance (CIR) effect \cite{olshanii98}. The narrow transversal confinement leads to a singularity
of the coupling constant for the corresponding effective longitudinal motion and as a result
we encounter a 1D strongly interacting gas of impenetrable bosons that can be mapped at resonance
on a gas of free fermions, the so-called Tonks-Girardeau gas \cite{girardeau60}.
The CIR has been interpreted as a Feshbach resonance
involving the atom-atom continuum and a bound state of a transversally excited channel of the harmonic waveguide
\cite{bergeman03,melezhik07,saeidian08}.
CIRs have subsequently been discovered for three-body \cite{mora04,mora05} and
four-body \cite{mora05a} scattering under confinement and for a pure p-wave scattering of fermions
\cite{granger04}. The CIR behaviour has been found
experimentally for s-wave scattering bosons \cite{kinoshita04,paredes04} and
for p-wave interacting fermions \cite{guenter05}. Recently a dual-CIR, which leads to a confinement-induced
transparency effect \cite{kim06,melezhik07}, has been demonstrated.
It is characterized by a complete suppression of strong
s- and p-wave heteronuclear atomic scattering in 3D because of the presence of the waveguide.

It is to be expected that waveguides with an anharmonic and/or anisotropic transversal confinement
will lead to novel ultracold atomic collision properties and consequently to an intriguing many-body dynamics
of the quantum gas. Very first results in this direction include the
observation of a modified CIR due to a coupling of the center of mass (CM) and relative motion
of the atoms \cite{peano05,melezhik07}. This nonseparability can be either achieved by
employing anharmonic waveguides or, in case of heteronuclear collisions, by the fact that
the two species experience different harmonic frequencies.

In the present work we show that the quantum dynamics of the coupled CM
and relative motion for two-atomic species feeling different confinement potentials, i.e.
waveguides, exhibits a resonant formation process of ultracold molecules. The latter is
analyzed in detail and its existence is found to be independent of the specifics of the atomic
interaction. As a consequence it is anticipated that heteronuclear atomic-molecular 1D quantum
gases in an anisotropic waveguide should explore a rich many-body quantum dynamics.

\section{Wave-packet propagation method}
We employ a wave-packet dynamical approach to study the atomic scattering in waveguides which
has been developed recently \cite{kim06,melezhik07}. The collisional dynamics of two (distinguishable) atoms
with coordinates $\ve{r}_1,\ve{r}_2$ and masses $m_1, m_2$ moving in the harmonic waveguide with
the transverse potential $\frac{1}{2} \sum_i m_i \omega_i^2 \rho_i^2$ ($\rho_i = r_i sin \theta_i$)
is described by the 4D time-dependent Schr\"odinger equation ($\hbar =1$)
$$
i\frac{\partial}{\partial t}\psi(\rho_R,\ve{r},t)=H(\rho_R,\ve{r})\psi(\rho_R,\ve{r},t)
$$
with the transformed Hamiltonian
\beq
H(\rho_R,\ve{r})= H_{\perp}(\rho_R) + H_{\mu} (\ve{r}) +
W(\rho_R,\ve{r})+V(r)\,\,.
\eeqn{1}
Here
\beq
H_{\perp}= -\frac{1}{2M}(\frac{\partial
^2}{\partial \rho_{R}^{2}}+\frac{1}{4\rho_{R}^{2}}) -
\frac{1}{2M\rho_{R}^{2}}
(\frac{\partial}{\partial \phi_R} -\frac{\partial}{\partial\phi})^{2}
+\frac{1}{2}(m_{1}\omega_{1}^{2}+m_{2}\omega_{2}^{2})\rho_{R}^{2}
\eeqn{2}
and
\beq
H_{\mu} = -\frac{1}{2\mu}\frac{\partial ^{2}}{\partial r^{2}}+
\frac{L^{2}(\theta,\phi)}{2\mu
r^{2}}+\frac{\mu^2}{2}(\frac{\omega_{1}^{2}}{m_{1}}+\frac{\omega_{2}^{2}}{m_{2}}) \rho^2
\eeqn{3}
describe the CM and relative atomic motions, $V(r)$ describes the atom-atom interaction, $\rho_R$ and $\ve{r} = \ve{r}_1 - \ve{r}_2
\mapsto (r, \theta, \phi) \mapsto (\rho, \phi, z)$ are the polar radial CM and the relative coordinates and
$M=m_1+m_2$, $\mu = m_1 m_2/ M$. The transformed Hamiltonian (\ref{1}) is
given in a rotated frame which exploits the conservation of the component
($L_1+L_2$) of the total angular momentum and allows to eliminate the
angular dependence of $\phi_R$ of the CM degrees of freedom
\cite{melezhik07,bock05}. Our investigation will focus on the case
$M_{\phi_R}=0$, the later being the quantum number belonging to $\frac{1}{i}\frac{\partial}{\partial \phi_R}$.

%In deriving the above Hamiltonian it was assumed that the conserved
%total angular momentum projection onto the waveguide axis vanishes.

The term
\beq
W(\rho_R,\ve{r})=\mu(\omega_{1}^2-\omega_{2}^{2}) r \rho_{R} \sin \theta \cos \phi
\eeqn{3a}
in the Hamiltonian (\ref{1}) leads for two distinguishable atoms that feel different confining
frequencies $\omega_1\neq\omega_2$ to a nonseparability of the CM and relative atomic
motion. We integrate the Schr\"odinger equation from time $t=0$ to the
asymptotic region $t\rightarrow +\infty$ with the initial wave-packet

\beq
\psi(\rho_{R},\ve{r},t=0)=N r \sqrt{\rho_R}
\exp\{-\frac{\rho_1^2}{2a_1^2} - \frac{\rho_2^2}{2a_2^2}
-\frac{(z-z_{0})^{2}}{2a_{z}^{2}}+ik_{0}z\}
\eeqn{4}
representing two different noninteracting atoms in the transversal ground state of the waveguide
with $a_i=(1/m_i\omega_i)^{1/2}$ and the overall normalization constant $N$
defined by $< \psi(0)|\psi(0)> = 1$. We choose
$z_0\rightarrow -\infty $ to be far from the origin $z=0$ and
$a_z\rightarrow\infty$ to obtain a narrow width in momentum and energy space for
the initial wave-packet.
Our wave-packet moves with a positive interatomic velocity $v_{0}=k_0/\mu=\sqrt{2\varepsilon_{\|}/\mu}$
thereby approaching the scattering region located at $z=0$.
In the course of the collision, the wave-packet splits up into two parts moving in opposite directions
$z\rightarrow \pm\infty$. We model the interatomic interaction $V(r)$ via the Lennard-Jones 6-12 potential
$V(r)=C_{12}/r^{12}-C_{6}/r^{6}$.
%\begin{equation}
%\label{yukawa}
%V(r)=V_0\frac{r_0}{r}\, e^{-r/r_0}
%\end{equation}
Note, that the effects discussed in the following exist for even qualitatively
very different shapes of the interatomic potential (like 6-12 and screened Coulomb \cite{melezhik07,saeidian08}).
In this sense, our results will be universal, although the molecular formation probabilities, to some
limited extend, depend on the appearance of $V(r)$. To be specific we consider the pair collisions of $^{40}$K and $^{87}$Rb atoms
with mass ration $m_1/m_2=40/87$ and the trapping frequency $\omega_2=2\pi\times 200$ kHz for Rb. Hereafter we use the units $\mu=\hbar=\omega_0=1$ with
$\omega_0=2\pi\times 10$ MHz. For the case of a decoupled CM motion, $\omega_1=\omega_2=0.02$ in these units, whereas for the coupled case
$\omega_1$ varies in the limits $\omega_2 < \omega_1 \le 2.2 \omega_2$.
Our focus is the analysis of the population dynamics of molecular
resonance states (for reasons of brevity we will simply refer to them as molecular bound states
with respect to $V(r)$) for ultracold atom-atom collision in the waveguide.

\section{Two-body bound states in a harmonic waveguide}

First we analyze how the molecular spectrum changes under
the action of the confining waveguide (see Fig.1).

We allow $C_{12}$ to vary for fixed $C_{6}=1.847$ \cite{derevyanko01} and focus on the
regime of the appearance of a bound state of $V(r)$.
In case of the absence of the confining potential the first bound molecular
state $\varepsilon_0<0$ appears at $C_{12}\simeq 0.13$ and becomes increasingly bound
with decreasing $C_{12}$. To explore the dependence of the binding energies of these states on $C_{12}$ in the presence of
the waveguide we solve the corresponding eigenvalue problem of the
four-dimensional Hamiltonian $H(\rho_R,\ve{r})$ (\ref{1}) employing the spectral method elaborated in ref.
\cite{FeitFleck}. It is based on the computation of the autocorrelation function
$<\psi(\rho_R,\ve{r},t=0)|\psi(\rho_R,\ve{r},t)>$ where the initial
state $\psi(\rho_R,\ve{r},t=0)$ can be considered as a test function of
the spectrum. The solution $\psi(\rho_R,\ve{r},t)$ of the time-dependent Schr\"odinger
equation is obtained via our wave-packet propagation
method \cite{melezhik07}. For the case $\omega_1=\omega_2=\omega$, i.e. for a decoupled
CM motion, every unconfined bound state with energy $\varepsilon_0(C_{12})$ transforms
by the action of the waveguide into the spectrum $\varepsilon_{n_1n_2}(C_{12})$ which represents at $C_{12} \geq 0.15$
\begin{equation}
\varepsilon_{n_1n_2}(C_{12})\rightarrow
(\omega_1+\omega_2)+2n_1\omega_1+2n_2\omega_2=2\omega(1+n_1+n_2)\,\,
\label{9}
\end{equation}
the spectrum of two independent identical two-dimensional oscillators
with radial quantum numbers $n_1=0,1,...$
and $n_2=0,1,...$ characterizing the transverse excitations of two noninteracting atoms in the waveguide.
We use $n_1,n_2$ as labels of the molecular bound states $\varepsilon_{n_1,n_2}(C_{12})$
at fixed $C_{12}$ correlating to the corresponding dissociation limit (\ref{9}) where they become exact
quantum numbers. It is evident from (\ref{9}) that for $\omega_1=\omega_2$ the excited
states $(n_1,n_2)$ become degenerate $\varepsilon_{n_1n_2}=\varepsilon_{n_2n_1}$.
Since the coupling between the CM and the relative motion is absent the
quantum numbers $n$ and $N$ of the harmonic
spectrum of the relative and the CM transverse eigenstates
could be equally used to classify the spectrum.

\begin{figure}[t]
\centering
\includegraphics[scale=0.76]{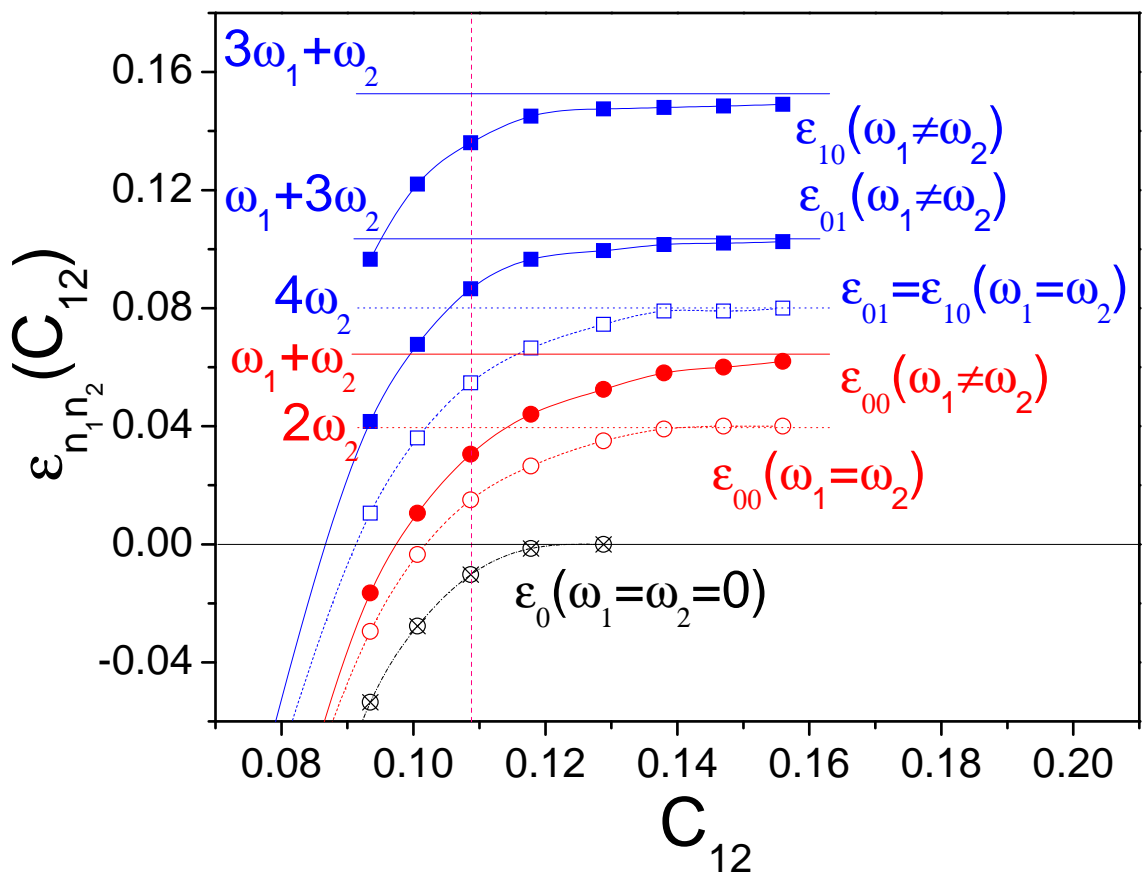}
\caption{ \label{fig1} (color online)
Molecular energy levels $\varepsilon_0(C_{12})$ and
$\varepsilon_{n_1,n_2}(C_{12})$ as a function of the parameter $C_{12}$ of the atomic interaction in free space
(open crossed circles) and in the waveguide. $\omega_1 /2.2 = \omega_2 =0.02$ for different
transversal frequencies and $\omega_1 = \omega_2 =0.02$ for the same frequencies.}
\end{figure}

For $\omega_1 \neq \omega_2$ the presence of the coupling between the CM and the relative
motion leads to a lifting of the above-mentioned degeneracy. In Fig.\ref{fig1} we show the
corresponding splitting $\varepsilon_{01} \neq \varepsilon_{10}$ of the first excited
state and the shift of the ground state.

\section{Molecular resonance states in waveguides}
The above-discussed spectral structure of the diatomic molecule in the waveguide
permits us to analyze the dynamics of atomic pair collisions with respect to the formation
of molecular final states. We choose herefore $C_{12}=0.109$ and the energy of the
colliding atoms to be $\varepsilon=\omega_1+\omega_2+\varepsilon_{\|}$
between the thresholds of the lowest $\omega_1+\omega_2$ and the first excited
$\omega_1+3\omega_2$ transverse channels ($0\leq\varepsilon_{\|}\leq 2\omega_2$).
Fig.1 indicates that for the chosen potential and collision energies one can observe one bound
state $\varepsilon_{01}$ of the closed channel $n_1=0,n_2=1$ which becomes
degenerate $\varepsilon_{01}=\varepsilon_{10}$ for $\omega_1=\omega_2$.
Fig.2 illustrates the time-evolution of the wave-packet in the course of the atomic collision
for the case of CM nonseparability $\omega_1\neq\omega_2$ (Fig.2a) and CM
separation $\omega_1=\omega_2$ (Fig.2b).

\begin{figure}[t]
\centering
\includegraphics[scale=.9]{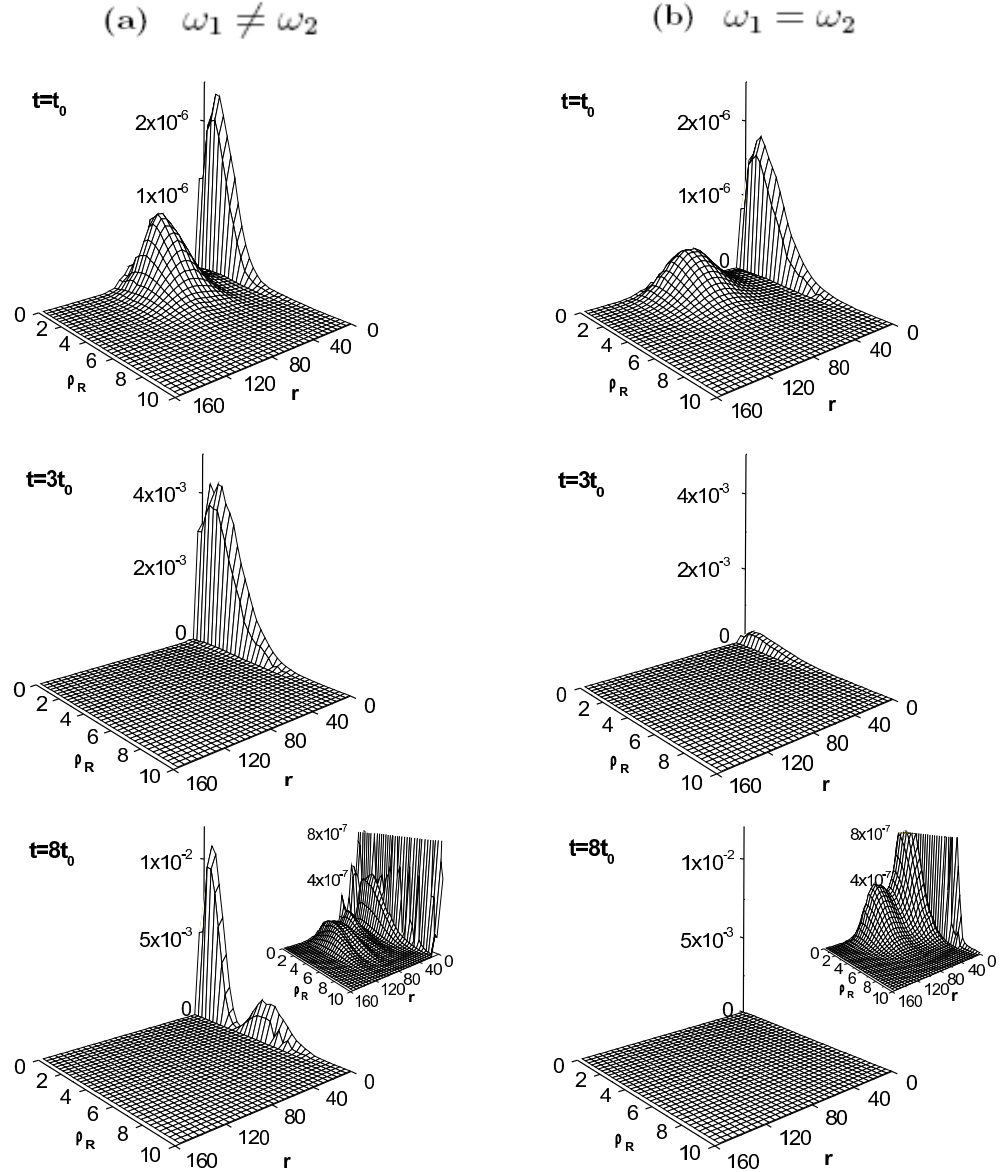}
\caption{\label{fig2} Time evolution of the probability density  distribution averaged over
the angular variables $W(\rho_R,r,t) = \int|\psi(\rho_R,r,\theta,\phi,t)|^2 (r^2\rho_R)^{-1} \sin\theta d\theta d\phi$.
({\bf a}) For the waveguide with $\omega_1/2.2=\omega_2=0.02$
({\bf b}) for $\omega_1=\omega_2=0.02$. For $t=8t_0$ corresponding to a time
after the collision the insets show a more detailed view of $W$ on the scales
$0\leq W \leq 8\times 10^{-7}$ corresponding to the continuum part of the wave-packet.
$\varepsilon_{\|}=0.004$, time is given in units of $t_0=\pi/\omega_2$.}
\end{figure}

For $\omega_1 \neq \omega_2$ we observe that a considerable part of the
scattered wave-packet (see the probability density distribution
$W(\rho_R,r,t)$ at $t=8t_0$) is located near $r=0$ after the collision and
corresponds to a molecular bound state $\varepsilon_{01}$
which in the $(n,N)$-representation is a mixture of the two states $n=0$,$N=1$ and $n=1$,$N=0$
with dominating $n=0$,$N=1$ contribution. Note that during collision
$t\sim3t_0$ the main part of the wave-packet is temporarily in the
ground state $\varepsilon_{00} (n=N=0)$, decays thereafter rapidly into the
continuum but part of it goes into the excited molecular state $\varepsilon_{01}$.
In contrast to this the case $\omega_1 = \omega_2$
in Fig.2b (CM separation) shows an almost complete decay into the continuum:
The remaining minor part near $r=0$ is much smaller compared to the case
$\omega_1 \ne \omega_2$.

To quantitatively investigate the molecular formation process
accompanied by the CM excitation we calculate the population probability
$P_N(t)$ of the molecular bound states for $N=0$ and $N=1$
\begin{equation}
P_N(t)= \int_0^{r_m} d r \int d\Omega~~|\int_0^{\infty}d \rho_R\psi(\rho_R,\ve{r},t)\Phi_N(\rho_R)|^2
\label{10}
\end{equation}
where $\Phi_N$ are the two-dimensional oscillator states of the potential
$(1/2) (m_1\omega_1^2+m_2\omega_2^2) \rho_R^2$ and $d\Omega=\sin\theta d\theta d\phi$. For the chosen interval of collision energies
$0\leq\varepsilon_{\|}\leq 2\omega_2$ and $C_{12}=0.109$ the molecular states $N=0$ and $N=1$ are the ones which dominate
during the collision process (see Fig.1). Note, that the integration in (\ref{10})
over the interatomic distance $r$ is limited to the region $r\leq r_m=10$
of the action of the 6-12 potential. For sufficiently long times, i.e. after the collision
process is concluded, all the unbound part of the scattered wave-packet resides outside this region. The
complementary part is given by $P_N(t)$ and can be interpreted as a molecular bound state.
For $r_m\rightarrow\infty$ we obtain $P_0(t)+P_1(t) \approx 1$ for any time and any energy within the
above-provided interval. Fig.3 shows ($\omega_1 \ne \omega_2$)
that during the time interval of close collision
the wave-packet temporarily occupies the molecular ground state $N=0$ i.e. $P_0$ becomes large.
However, with further increasing time $P_0$ decays to zero and $P_1$ increases
rapidly which corresponds to the population of the excited state $N=1$. $P_1$ decays very slowly
for long enough times after the collision. We interprete this state
as the first excited state $n_1=0, n_2=1$ of the molecular spectrum (see Fig.1).
For $\omega_1=\omega_2$ the coupling between the $n_1=n_2=0$ ($N=0$) and $n_1=0, n_2=1 (N=1)$ molecular states is absent and
the temporary population in the course of the collision of the molecular ground state
decays back into the continuum after the collision (see behaviour of $P_0(\omega_1=\omega_2)$.
In particular $P_1(\omega_1=\omega_2)$ is negligible always.

\begin{figure}[t]
\centering
\includegraphics[scale=0.75]{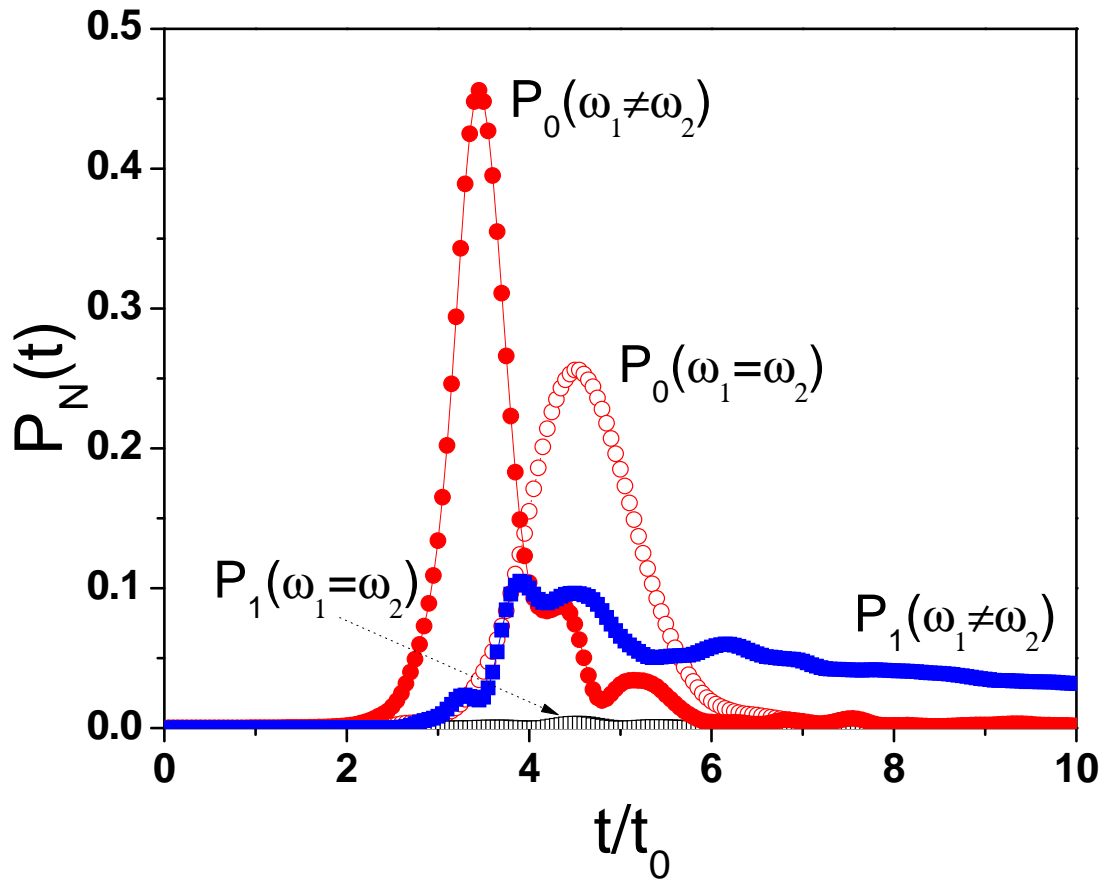}
\caption{ \label{fig3}
(color online) $P_N(t)$ of the states $N=0,~1$ as a function of the collision time $t$.
Full circles and squares correspond to $\omega_1/2.2=\omega_2=0.02$. Open circles and squares
belong to $\omega_1=\omega_2=0.02$. $\varepsilon_{\|}=0,004$, time is given in units of $t_0=\pi/\omega_2$.}
\end{figure}

\begin{figure}[t]
\centering
\includegraphics[scale=0.75]{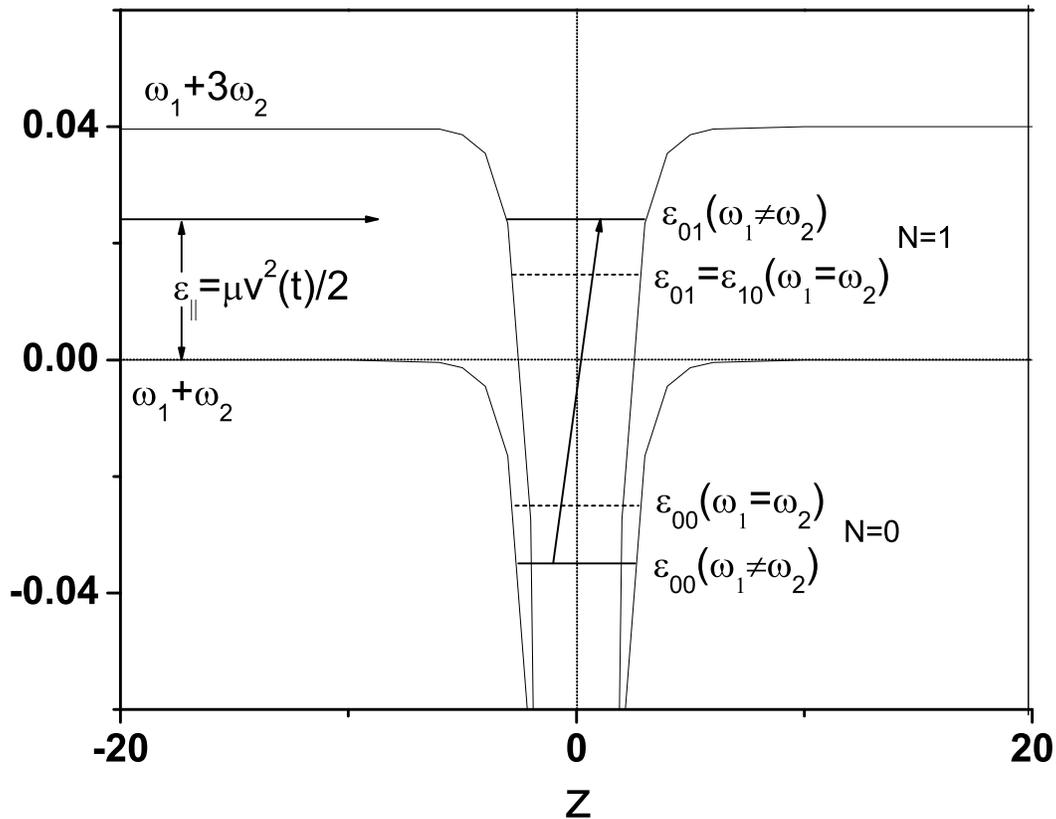}
\caption{ \label{fig4}
Schematic illustration of the resonant formation of a diatomic molecule.
Transfer of the released energy $\varepsilon_{\|}+(\omega_1+\omega_2)-\varepsilon_{00}$ to the
excitation $\varepsilon_{01}-\varepsilon_{00}$
of the CM in the course of molecular formation. Solid lines indicate the positions of the molecular
energy levels $\varepsilon_{n_1,n_2}$ for the case $\omega_1/2.2=\omega_2=0.02$.
The shifts of the levels due to the elimination of the coupling term at $\omega_1=\omega_2=0.02$
are shown by dashed lines. }
\end{figure}

\begin{figure}[t]
\centering
\includegraphics[scale=0.75]{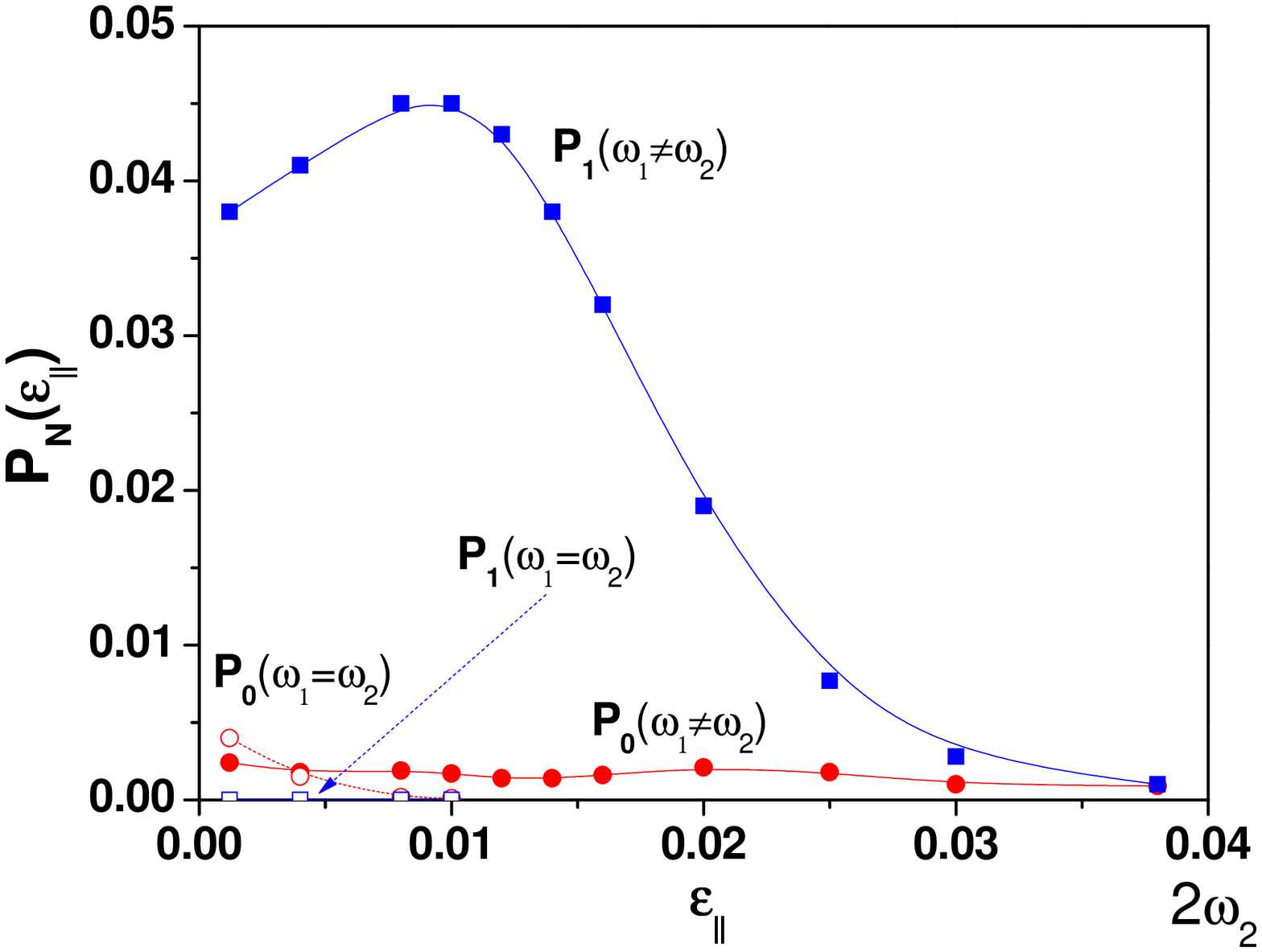}
\caption{ \label{fig5}
(color online) $P_N(\varepsilon_{\|})$ for the states $N=0$ and $1$
as a function of the longitudinal energy $\varepsilon_{\|}$ before the collision. Full
circles and squares correspond to $\omega_1/2.2=\omega_2=0.02$.
Open circles and squares correspond to $\omega_1=\omega_2=0.02$.}
\end{figure}

In Fig.\ref{fig4} we illustrate the mechanism of molecule formation for
the ground state $n_1=n_2=0 (N=0)$ with transversal $\omega_1+\omega_2$ and relative longitudinal
$\varepsilon_{\|}=\mu v_0^2/2$ energies. Under the action of the attractive tail
$-C_6/r^{6}$ of interatomic interaction the atoms accelerate up to the energy $\mu
v^2(t)/2=\varepsilon_{\|}+\omega_1+\omega_2-\varepsilon_{00}$ at the instant of collision.
The occurence of the molecular state with energy $\varepsilon_{00}$
(bound molecular state with respect to the ground transversal channel)
as an intermediate in the course of the collision process can be seen by inspecting the
intermediate probability density shown in Fig.2a and Fig.3.
The released energy subsequently transfers to the closed channel $n_1=0,n_2=1 (N=1)$ via formation
of a molecule in a CM excited state since this closed channel is coupled to the $N=0$  by the term
$W(\rho_R,\ve{r})=\mu (\omega_1^2-\omega_2^2) r \rho_R \sin\theta \cos\phi$ (\ref{3a}) of the total
Hamiltonian (\ref{1}). This mechanism also explains the delay of the onset of the collisional
interaction observed in Fig.\ref{fig3} in the
absence of the coupling term $\omega_1 = \omega_2$ compared to the case $\omega_1\neq\omega_2$.
For $\omega_1 = \omega_2$ the collisional interaction happens later than for
$\omega_1\neq\omega_2$ because the binding energy in the channel $\varepsilon_{00}
(\omega_1\neq\omega_2)$ is considerably lower than the binding energy in the channel $\varepsilon_{00}
(\omega_1 = \omega_2)$ leading to a stronger atomic attraction and
acceleration of the colliding atoms for $\omega_1\neq\omega_2$. A simple semiclassical estimate of
the time difference between the onsets of collisional interactions
for the two cases is in good agreement with the numerical result of Fig.\ref{fig3}.

From the above scheme it also follows that if the energy release
$\varepsilon_{\|} +\omega_1+\omega_2-\varepsilon_{00}$ is equal to the excitation energy
$\varepsilon_{01}-\varepsilon_{00}$ we have to expect a resonant enhancement of the
molecular formation process i.e. we encounter the resonance condition
\begin{equation}
\varepsilon_{\|} +\omega_1+\omega_2 = \varepsilon_{01} \,\,\,
\label{11}
\end{equation}
To analyze the energy dependence of the atom-molecule reaction we inspect the
population probabilities $P_N(\varepsilon_{\|})$ for the molecular
states $N=0$ and 1 as a function of the longitudinal energies $\varepsilon_{\|}$ (see Fig.\ref{fig5}).
The values $P_N(\varepsilon_{\|})=P_N(\varepsilon_{\|},t\rightarrow\infty)$ were
calculated after the collision at $t=8t_0$, in the asymptotic region where
$P_1(\varepsilon_{\|},t)$ is very slowly decaying if $\omega_1\neq\omega_2$.
$P_1(\varepsilon_{\|})$ exhibits a pronounced resonant behaviour whereas
$P_0(\varepsilon_{\|})$ shows a very weak energy dependence which is in agreement
with the above-discussed reaction mechanism. The calculated position
$\varepsilon_{\|}=\varepsilon_r \sim 0.01$ of the maximum of
the population $P_1(\varepsilon_{\|})$ is shifted considerably compared to
$\varepsilon_r(\omega_1\neq\omega_2)=\varepsilon_{01}(\omega_1\neq\omega_2)-\omega_1-\omega_2=0.026$ following
from the resonance condition (\ref{11}). We attribute this difference to the fact that we are
using for the calculation of $P_1(\varepsilon_{\|})$
in eq.(\ref{10}) the oscillator wave function $\Phi_N(\rho_R)$ instead of the
exact transversal CM part of the molecular wave function $n_1=0, n_2=1 (N=1)$.
The resulting deviation from the exact resonance conditions (\ref{11}) is particularly due to
the missing effects of the CM coupling which provides
the shift $\varepsilon_{01}(\omega_1=\omega_2)- \varepsilon_{01}(\omega_1\neq\omega_2)=0.014$
resulting in a significantly better agreement of the calculated position of the resonance with the
resonant energy from eq.(\ref{11}).

The molecular formation probability $P_1$ depends also on $\Delta \omega = \omega_1 - \omega_2$ and
reaches a maximum at a value corresponding to the resonance condition (\ref{11}).
This probability can therefore be controlled by
changing the transversal confinement frequencies for the different atomic species.
By nature the process we observed is a molecular resonance in the atomic scattering and back decay
of the molecule to the atom-atom continuum will occur. However, this resonance possesses a long lifetime
due to its two-step character which is clearly demonstrated in Fig.\ref{fig3}.
Finally we note that $P_1$ amounting to several percent should not obscure
the fact that substantial molecular formation rates $\lambda=n_A v_0 P_1$ could be achieved by integrating
over many collisions even for low linear atomic densities $n_A$. Thus, for
$n_A=10^4$cm$^{-1}$ and $T=100$nK we obtain the estimate $\lambda\sim
10^3$s$^{-1}$.

\section{Conclusions}
Perspectives concerning applications of the molecular formation mechanism are based on the fact that it
represents a key ingredient for the preparation and following investigation of the dynamics
of mixed atomic and dipolar molecular quantum gases in waveguides. The latter are expected to
possess a novel collective excitation dynamics particularly in the ultracold
regime discussed here for which only a few transversal channels
are accessible. If wanted, one could stabilize the molecular gas and prevent back decay
into atoms by a corresponding switch of the trapping fields such that the CM decouples
from the interatomic motion finally. Open questions include the impact of inelastic atom-molecule
and molecule-molecule collisions, which however goes beyond the scope of the present investigation.

\section*{Acknowledgments}
P.S. gratefully acknowledges financial support by the Deutsche Forschungsgemeinschaft.
Financial support by the Heisenberg-Landau Program is appreciated by P.S. and
V.S.M. We thank the Heidelberg Graduate School of Fundamental Physics for continuing support.

\end{document}